\documentclass[letters,fleqn,usenatbib,useAMS]{mnras}

\usepackage{graphicx}	
\usepackage{amsmath}	
\usepackage{amssymb}	
\usepackage{multicol}        
\usepackage{bm}		
\usepackage{pdflscape}	

\def\Planck{\textit{Planck}}

\newcommand{\vnvec}{\mathbf{\hat{n}}}

\newcommand{\dd}{\mathbf{d}}

\newcommand{\hMpc}{ h^{-1}{\rm Mpc}}

\def\apj{{ApJ}}                 
\def\apjs{{ApJS}}               
\def\aap{{A\&A}}                

\def\jcap{{J. Cosmology Astropart. Phys.}}
\def\mnras{MNRAS}             
\def\prd{{Phys.~Rev.~D}}        
\usepackage[T1]{fontenc}
\usepackage{ae,aecompl}

\usepackage{newtxtext,newtxmath}

\title[ARF gravity constraints in the late universe]{Tomographic Constraints on Gravity from Angular Redshift Fluctuations in the Late Universe}
\author[C.Hern\'andez-Monteagudo et al.]{Carlos~Hern\'andez--Monteagudo,$^{1}$
\thanks{Recently moved to Instituto de Astrof\'\i sica de Canarias (IAC), Tenerife, Spain. Contact e-mail address: \href{mailto:chm@cefca.es}{chm@iac.es}}
Jon\'as Chaves-Montero,$^{2}$
Ra\'ul E. Angulo,$^{3,4}$
\newauthor
and Giovanni Aric\`o.$^{3,5}$
\\
$^{1}$Centro de Estudios de F\'\i sica del Cosmos de Arag\'on (CEFCA), Unidad Asociada al CSIC, Plaza San Juan, 1, planta 2, E-44001,Teruel, Spain\\
$^{2}$HEP Division, Argonne National Laboratory, 9700 South Cass Avenue, Lemont, IL 60439, USA\\
$^{3}$Donostia International Physics Centre (DIPC), Paseo Manuel de Lardizabal 4, 20018 Donostia-San Sebastian, Spain.\\
$^{4}$IKERBASQUE, Basque Foundation for Science, E-48013 Bilbao, Spain.\\
$^{5}$Centro de Estudios de F\'\i sica del Cosmos de Arag\'on (CEFCA), Plaza San Juan, 1, planta 2, E-44001,Teruel, Spain}
\date{\today}

\pubyear{2020}

\begin{document}
\label{firstpage}
\pagerange{\pageref{firstpage}--\pageref{lastpage}}
\maketitle

\begin{abstract}
Fluctuations in sky maps of the galaxy redshifts, dubbed as angular redshift fluctuations (ARF), contain precise information about the growth rate of structures and the nature of gravity in the Universe. Specifically, ARF constrain the combination of cosmological parameters $H/H_0\,f\sigma_8(z)$, while being an intrinsically tomographic probe and largely insensitive to many observational systematic errors, all this without requiring the assumption of any redshift-to-distance relation under a given fiducial cosmology. We present the first cosmological constraints derived from ARF by using BOSS LOWZ+CMASS DR12 galaxy samples, obtaining 7\%-accurate constraints on $H/H_0 f\sigma_8(z)$ at more than 20 redshifts over the range $z \in [0.26,0.72]$. Our best-fitting value is $10\%$ larger, but compatible at the $1.4\sigma$ level, than the $\Lambda$CDM expectation set by {\it Planck} observations of the Cosmic Microwave Background (CMB) radiation. Our tomographic measurements, combined with these CMB data, provides one of the strongest constraints on the gravity index $\gamma$, $\gamma=0.44^{+0.09}_{-0.07}$, which lies within $2\sigma$ from the prediction of General Relativity ($\gamma_{\rm GR}\simeq 0.55$).
\end{abstract}


{\it Introduction.} The last two decades have witnessed an impressive progress in the field of Cosmology: analyses of the temperature and polarization of the Cosmic Microwave Background (CMB) radiation, combined with the study of the spatial distribution of galaxies and quasars, have provided a largely consistent physical picture of our Universe. However, the nature of the gravitationally-dominant component, dark matter, remains elusive as well as the source for the accelerated expansion, dubbed as ``dark energy" \citep{planck_parameters_15,Alam_2017,Beutler2017a,Beutler2017b,Chuang_2017,DES1yr_GC_WL,DES1yr_shear}. 

Indeed, dark matter and dark energy are two critical items in the list of open problems in Cosmology. Other important topics include the understanding of the early inflationary epoch; the testing of gravity on the largest possible scales; and the role of relativistic, light particles in the growth of structure in the Universe. For this purpose, ever more ambitious CMB and Large Scale Structure (LSS) experiments are under way \citep{actpol_16,simons_obs_18,Euclid_scaramella14,desi_2019,sphereX}, attempting to survey the observable Universe in great detail -- up to its largest scales and earliest epochs -- with tight control on all sources of observational error. 

The current interpretation of LSS measurements in terms of fundamental physics presents several limitations. First, cosmological analyses mostly focus on large scales, where the information can be more easily extracted. This is because those scales are close to the linear regime, where the complicated physics of galaxy formation and nonlinear evolution can be modelled more accurately. Unfortunately, these scales are also the most affected by observational errors (associated to, e.g. Milky Way star-background light, Galactic extinction, seeing, and other observational artifacts \cite[e.g.,][]{Ross_2011,Ross_2012,shafer2015,Laurent17}). In addition, current analyses typically need to assume an underlying cosmology when relating redshifts to comoving distances, which might add complications when testing departures from that fiducial cosmology. Finally, these analyses are performed over broad redshift intervals, which might hide peculiar features of the expansion and growth history of the Universe.



In this work we focus on a new LSS statistic -- angular redshift fluctuations (ARF) -- recently introduced by \citep{arf_lett1} which features several improvements with respect to traditional 3D LSS analyses. Using ARF and the SDSS DR12 spectroscopic LOWZ and CMASS galaxy samples \citep{sdss_III,cuesta_bao_dr12}, we provide the {\em first tomographic} measurements of the parameter combination $H(z)/H(z=0)\,f\,\sigma_8 (z)$, that is intimately related to the universal growth of structure, for more than 20 redshift bins over $z\in [0.26,0.72]$. 
We emphasize that our LSS statistic does not make any assumptions about the cosmological model in the measurements unlike standard 3D clustering estimators, while behaving more robustly wrt systematics biasing the observed number of probes\footnote{ARF estimators used in \citet{arf_lett1,Chaves-Montero_kSZ} show different degrees of robustness against systematics biasing the observed number of tracers, motivating an ongoing study on an optimal ARF estimator definition. Those works proved that such robustness lies in the ARF sensitivity to radial gradients of density/velocity under the redshift shells, that is (almost) blind to systematics. },
and using mostly linear scales in the density and velocity fields. In passing, we note that standard angular density fluctuations (ADF) share with ARF the properties of not assuming any fiducial cosmological model and being sensitive to $H(z)/H(z=0)f\sigma_8(z)$, although ADF are more subject to biases in the presence of systematics and non-linear physics, as it will be shown below.


\begin{figure*}
\includegraphics[width=0.4\paperwidth]{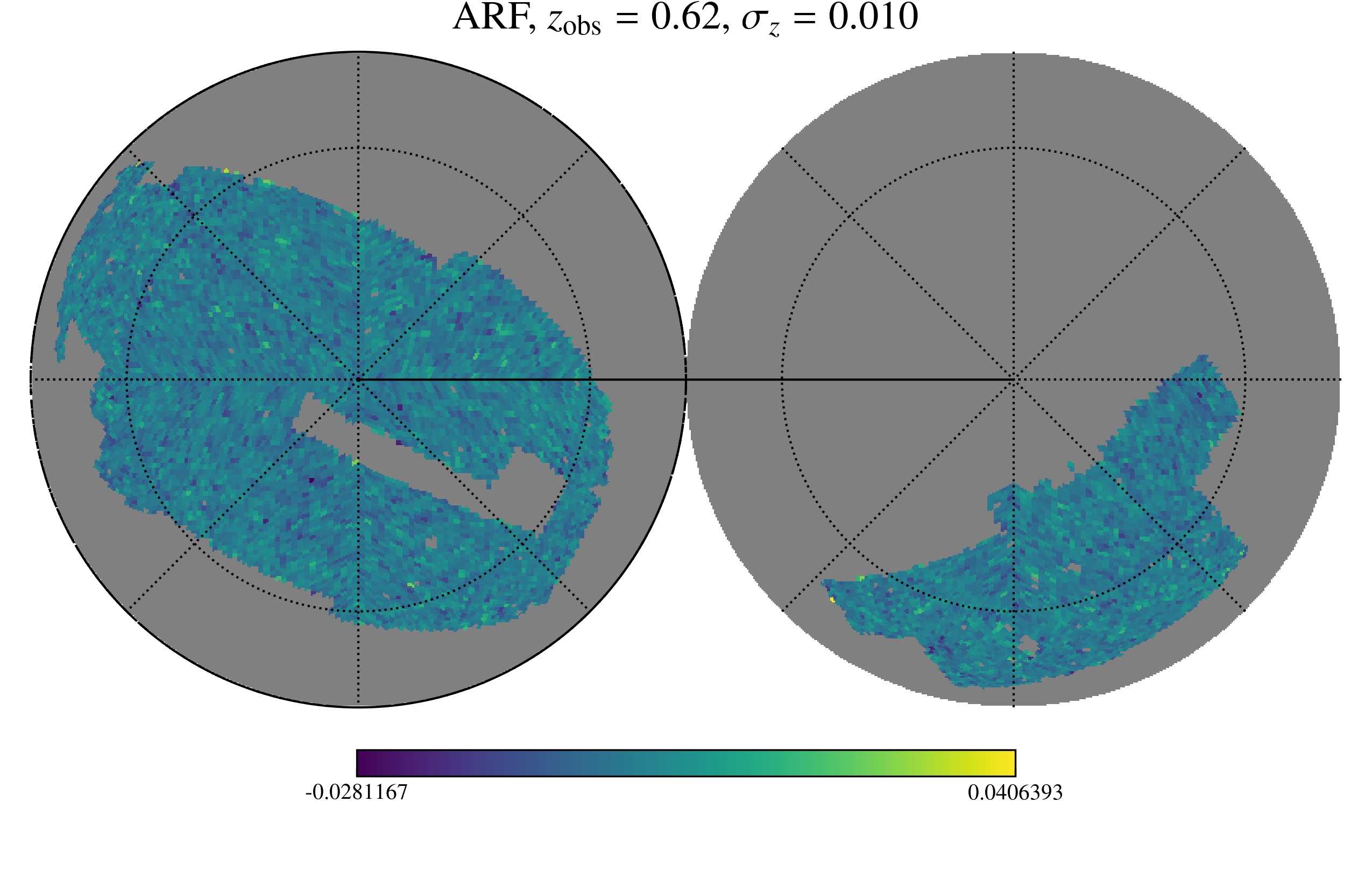}
\includegraphics[width=0.4\paperwidth]{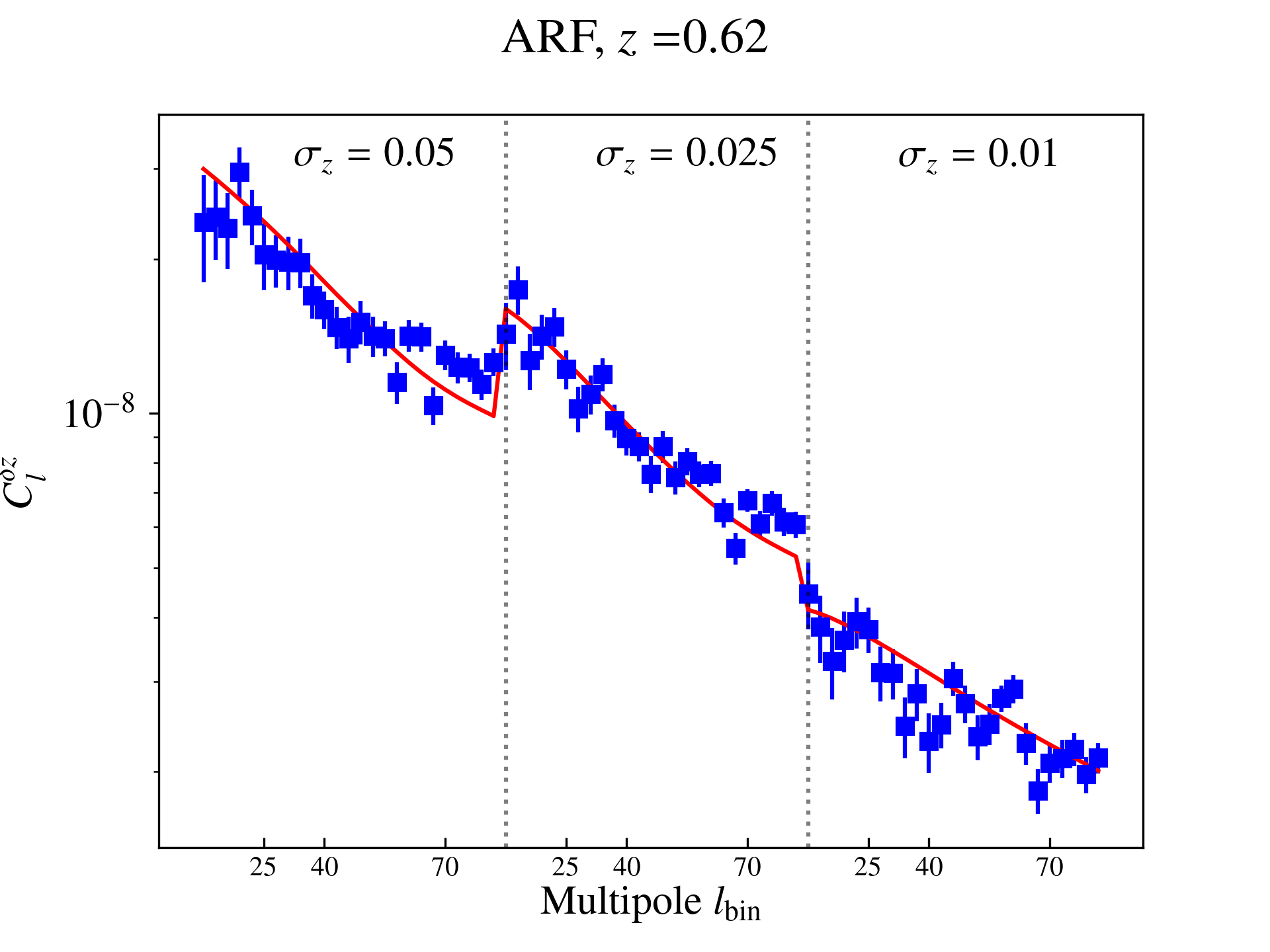}
\caption{ {\it Left panel:} Map of Angular Redshift Fluctuations (ARF) for CMASS galaxies under a Gaussian window centred on $z=0.62$ and width $\sigma_z=0.01$
{\it Right panel:} Measured ARF band angular power spectra for CMASS galaxies under a Gaussian shell centred on $z=0.62$, of widths $\sigma_z=0.05,\,0.025,\,$ and $0.01$. The best fitting model is displayed as a red solid line. }
\label{fig:F1}
\end{figure*}


{\it Methodology.} Our analysis is based on the SDSS-BOSS Data Release 12 (DR12) LOWZ and CMASS spectroscopic galaxy samples, which represent the most accurate LSS dataset to date and has been extensively characterized \citep{Ross_2012}. We analyze these galaxy samples in 20 redshift bins, centred upon $z_{\rm cen}$, separated by $\Delta z=0.02$ over the range $z_{\rm cen}\in [0.26,0.72]$. In each of these redshift intervals we consider galaxies with a Gaussian weight $W=\mathcal{G}(z_{\rm cen}, \sigma_z)$, where $\sigma_z=\{0.05, 0.025, 0.01\}$. Hence, we obtain 60 different (but overlapping) catalogues. 

For each sample we compute ADF and ARF as: $\delta_g(\vnvec) = \sum_{j\in \vnvec} W_j / \langle W_i\rangle-1$ and $\delta_z(\vnvec) = \sum_{j\in \vnvec} (z_j-\bar{z})W_j / \langle W_i\rangle$, respectively. The sum $\sum_{j\in \vnvec}$ selects galaxies along the sky direction $\vnvec$, and the Gaussian factor $W_j \equiv W(z_j)$ down-weights galaxies at $z_j$ far from the central redshift $z_{\rm cen}$. The average redshift under a given shell is $\bar{z}=\sum_{j} z_j W_j / \sum_j W_j$, where the sum runs over all galaxies, regardless of their angular location. Finally, the angular average of the weighted number of galaxies under the Gaussian shell over the entire survey footprint is $\langle W_i\rangle \equiv \sum_{\vnvec} \sum_{j\in \vnvec} W_j / \sum_{\vnvec} 1$, with the double sum over sky pixels/directions $\vnvec$ and galaxies falling within. 

We project the SDSS-BOSS sky footprint \footnote{ {\tt https://data.sdss.org/sas/dr12/boss/lss/}} into a HEALPix~\footnote{{\tt http://www.eso.org/science/healpix/} } mask using the code {\tt mangle}~\footnote{{\tt https://space.mit.edu/~molly/mangle/}, \cite{mangle}}. Given our typical number density of galaxies, we choose a HEALPix resolution parameter $N_{\rm side}=64$, which corresponds to $\sim1\deg^2$ pixels. We decompose $\delta_g(\vnvec)$ and $\delta_z(\vnvec)$ into spherical harmonics, $\delta_X (\vnvec) = \sum_{l,m} a^X_{l,m}Y_{l,m} (\vnvec)$, 
and then estimate the angular power spectrum as $C_l^X = \sum_{m=-l,l} |a^X_{l,m}|^2/(2l+1)$, where $X={g,\, z}$. The partial sky coverage of the LOWZ and CMASS galaxy samples  ($f_{\rm sky}\simeq 22\%$) biases low the measured angular power spectra, which we account for by convolving the theoretical power spectra with the sky mask \citep{master}. In addition, partial sky coverage induces correlation among low multipoles $\ell$, thus we average 3 consecutive multipoles in the measured spectra and consider only those centred on $l_{\rm bin}=10$ and above, i.e. we employ $l_{\rm bin}=10,13,16, ..., 121$. Finally, we correct the observed spectra by the effect of a finite HEALPix pixel size.

We highlight that we obtain almost identical ARF power spectra regardless of the use or not of galaxy weights provided by the BOSS collaboration, which attempt to mitigate the effect of observational systematic errors in large-scale clustering statistics. This insensitivity of ARF to systematics is consistent with the arguments provided in \cite{arf_lett1} and the tests conducted in \cite{Chaves-Montero_kSZ}.

\begin{figure*}
\includegraphics[width=0.4\paperwidth]{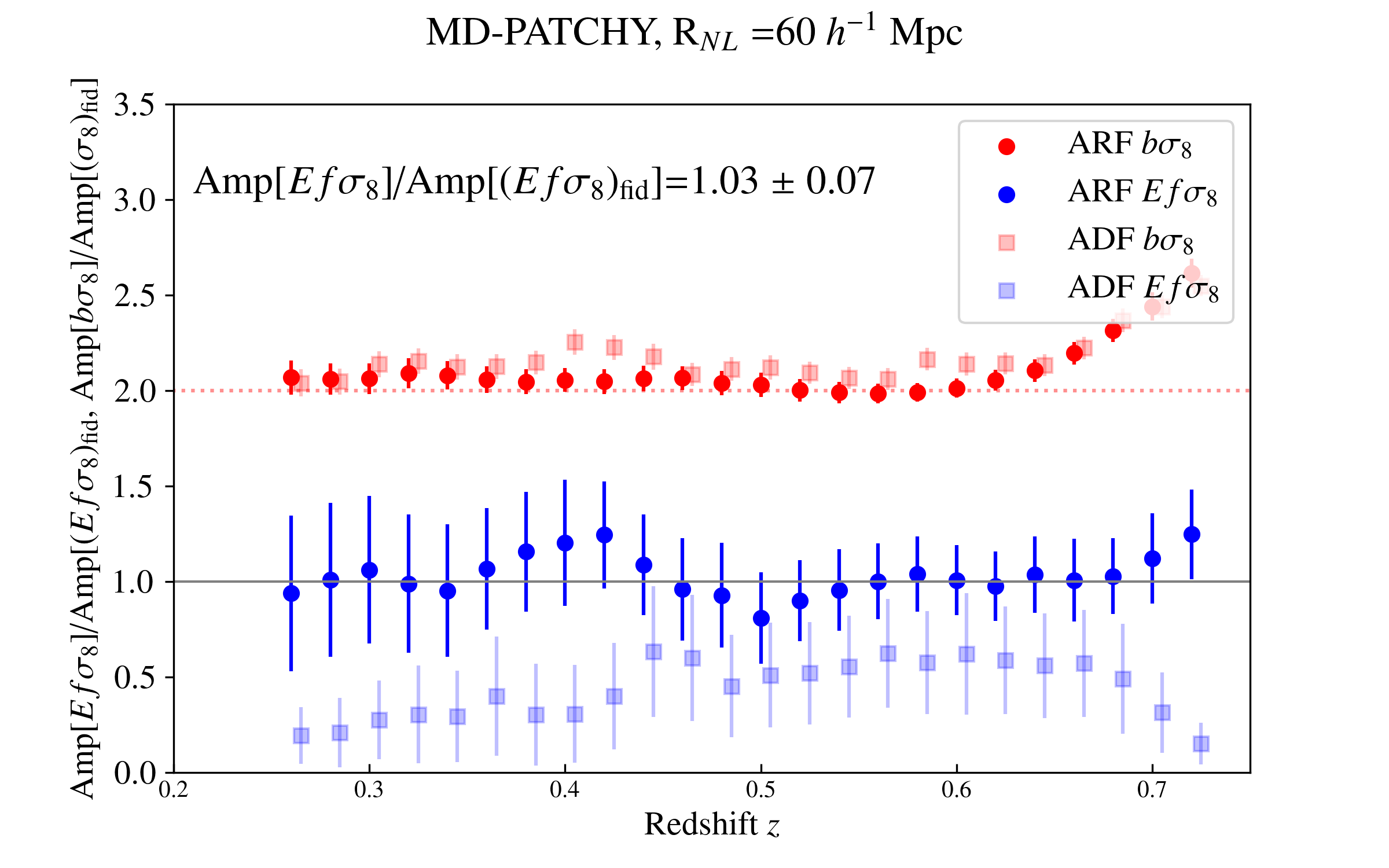}
\includegraphics[width=0.4\paperwidth]{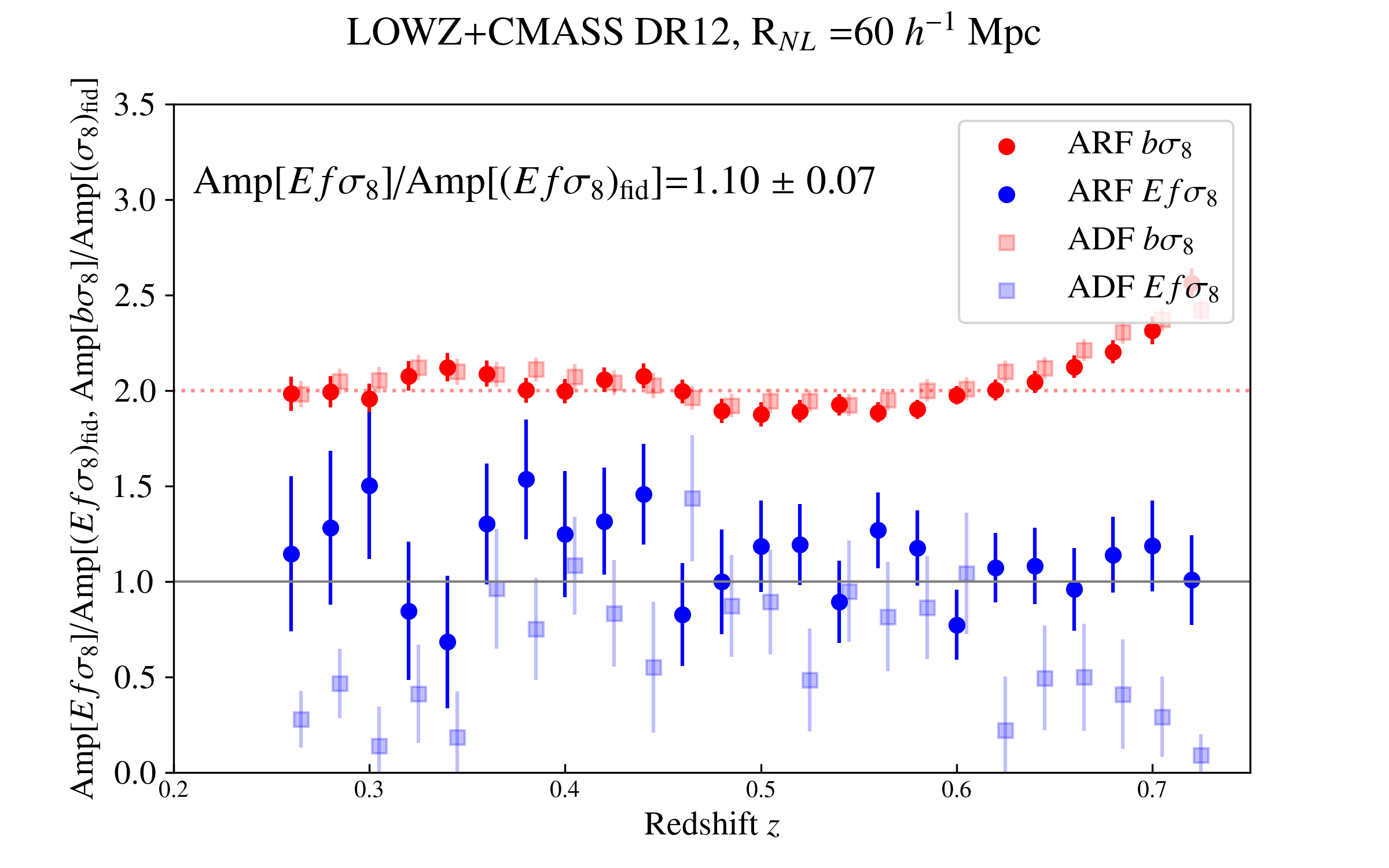}
\caption{Constraints on parameter combinations $b\sigma_8$ and $Ef\sigma_8$ in $20$ redshift bins. Left and right panels show results for 300 MD-PATCHY mocks and for LOWZ+CMASS DR12 galaxies, respectively. We show the ratio of the values obtained from our analysis pipeline over those in the {\it Planck} 2018 $\Lambda$CDM fiducial model, and adopting a fiducial large-scale bias $b_{\rm fid}=1$. Heavy symbols indicate measurements obtained from Angular Redshift Fluctuations (ARF) with $R_{\rm NL}=60~h^{-1}$~Mpc, whereas light symbols do so for measurements from standard Angular Density Fluctuations (ADF). Best-fitting values for $Ef\sigma_8$ from ARF measurements are quoted in the panels. }
\label{fig:F2}
\end{figure*}

We model our measurements with linear perturbation theory. Specifically, we employ the following model: 

\begin{eqnarray}
\label{eq:model}
C_l^X &=& (b_g\sigma_8)^2 C_l^{X,\delta\delta} + (Ef\sigma_8)^2 C_l^{X,vv} \nonumber \\
        &+& 2 (b_g\sigma_8)(E\,f\sigma_8)C_l^{X,\delta v} + C_l^{\rm SN},
\end{eqnarray}

\noindent where the $\delta$ and $v$ superscripts denote the auto or cross-correlation between density and velocity fields; $\sigma_8$ is the rms of the linear theory matter density fluctuations at present; 
$b_g$ is the bias of the observed galaxy clustering to that of matter; $E(z)\equiv H(z)/H(z=0)=H(z)/H_0$ where $H(z)$ is the Hubble parameter; $f$ is the growth rate function $f(z)\equiv d\log D_{\delta}/d\log a$; $D_{\delta}(z)$ is the linear growth factor of matter density fluctuations; and $a(z)=1/(1+z)$ is the cosmological scale factor. We note that the combination $E\,f\,\sigma_8$ captures the amplitude of radial, proper peculiar velocities. Additionally, $C_l^{\rm SN}$ describes the contribution of discreteness noise associated to the finite number of galaxies under analysis; and the parameter $\sigma_v$ (given in units of the speed of light $c$) accounts for small-scale random motions. It is present in the kernel ${\cal S}(k,\sigma_v) \equiv \exp{\bigl(-k^2/3\times [cH^{-1}(z)(1+z)\sigma_v]^2/2 \bigr)}$ that enters linearly in the $k$-integral defining $C_l^{X,\delta v}$, and squared in the same integral defining $C_l^{X,vv}$ \citep{arf_lett1}. We note that \citet{arf_lett1} showed that this model accurately reproduces the outputs of dark matter cosmological numerical simulations at the few percent level. 

For each central redshift $z_{\rm cen}$ we have a data vector given by $\dd=\{\dd_w,\,w=1,2,3 \}$ and $\dd_w=\left[ C^{g}_{l_{\rm bin}},\,C^{z}_{l_{\rm bin}} \right]_w$, where $w$ runs over our three widths ($\sigma_z=0.05,\, 0.025$, and $0.01$). These data are confronted to our theoretical model consisting of three physical parameters ($b_{\rm g}\sigma_8$, $Ef\sigma_8$, $\sigma_v$), and six amplitudes of $C^{\rm SN}_{l_{\rm bin}}$ (three shells for each ADF, ARF data vector). The shape of $C_l^{\delta \delta}$, $C_l^{\delta v}$, and $C_l^{vv}$ are extremely insensitive to the magnitude of variations of the cosmological parameters explored in the analysis (see below), thus, we have kept them fixed as dictated by the {\it Planck} best-fitting cosmology.

The impact of non-linear physics on small scales is minimized by restricting to measurements at $l < \pi/\theta_{\rm NL}(z)$, where $\theta_{\rm NL} (z_{\rm cen})$ is the angle subtended by a given physical scale $R_{\rm NL}$ at redshift $z_{\rm cen}$. We obtain unbiased results for $R_{\rm NL}$ values typically close to $60~h^{-1}$\,Mpc (see Table~\ref{tab:tab1}). 

We use a Monte-Carlo Markov Chain (MCMC) algorithm to measure the parameters of our model from our data vector. We assume a Gaussian likelihood with a covariance matrix estimated by measuring ARF and ADF spectra on a set of simulated galaxy mocks. We use two mock sets: a set of 100 COLA \cite{colaSIMs} mocks simulating the distribution of matter in the universe, and 300 MD-PATCHY mocks \citep{patchy}. These latter mocks were built to mimic the angular and radial selection function of BOSS LOWZ and CMASS galaxies, and constitute a standard analysis tool for those data. We also use the chains provided by the {\it Planck} collaboration in the context of the basic $\Lambda$CDM scenario, having typically few thousand steps. 
Note that, in principle, ARF measurements can be combined with ADF, but for reasons outlined below we will only consider ARF to infer cosmological constraints.

We find that the covariance matrix for the angular power spectra of ARF and ADF in the MD-PATCHY mocks differ significantly: while this matrix for ARF is practically diagonal, for ADF it shows significant off-diagonal structure (at the $10$--$50$~\% level) for multipole bins above $l_{\rm bin}>40$~\footnote{ADF in LOWZ+CMASS samples do contain, nevertheless, cosmological information, as shown in \cite{bossdr12_cl_tomography}.}. The reason for this difference is related to the distinct nature of the two probes: ARF are sensitive to (relatively small) redshift fluctuations with respect to a local average which are partially blurred/randomized by non-linear physics, whereas ADF reflect the scale coupling induced by non-linear gravitational evolution of density perturbations.



{\it Results.}  In Fig.~\ref{fig:F1} we show an example of an ARF map at $z_{\rm cen}=0.62$ and its corresponding band angular power spectrum for $\sigma_z=0.05$, $0.025$, and $0.01$ (blue squares, right panel).  We display the best fitting model (Eq.~\ref{eq:model}) as a red solid line. Note that the ARF map shows Gaussian, symmetric fluctuations (unlike an ADF map) and that the best-fitting model is an accurate description of the data, even on the lowest mutipoles, usually more affected by systematics. This serves as an example of the low degree to which observational systematics errors affect our ARF measurements.

We next address the constraints on the two parameter combinations that our method is sensitive to: $b\sigma_8$ and $E\,f\sigma_8$, after marginalising over all the other free parameters of our model. 
We first verify that our pipeline returns unbiased estimates of these parameters in the total matter, gravity-only COLA runs. As expected, for these mocks, retrieved values of $b\sigma_8$ are close to $\sigma_{8,\,{\rm fid}}=0.83$ (implying $b\simeq 1$), while $Ef\sigma_8$ also scatters around its fiducial expectation. We find that ADF and ARF contain complementary information on these two parameters, which is found to be compatible to each other, but impacted differently (at $\sim 7$--$15$\%) by the lack of galaxies in the two highest redshift bins, (see Table~\ref{tab:tab1}).
Next we examine in detail our pipeline outputs for the 300 MD-PATCHY mocks (left panel of Fig.~\ref{fig:F2}), and compare those to the outputs obtained from BOSS DR12 data (right panel of the same figure). 
In both cases we display values relative to the expectation of the reference cosmology employed in the galaxy mocks (given by {\it Planck}'s 2018 CMB measurements) and assuming a unity fiducial bias, i.e. $b_{\rm fid}=1$. Circles and squares indicate measurements obtained from ARF and ADF, respectively. For each case, red and blue symbols correspond to $b\sigma_8$ and $Ef\sigma_8$ amplitude estimates for each redshift bin. Error bars are given as the range comprising 68\% of the distribution of results from the MD-PATCHY mocks.

The left panel shows that our analysis pipeline provides statistically unbiased constraints from ARF for both parameters: ARF measurements on $Ef\sigma_8$ scatter around the unity, returning a value of $1.03 \pm 0.07$ for $R_{\rm NL}=60~h^{-1}$~Mpc, that is, recovered values remain unbiased at the $0.5 \sigma$ level. All redshift bins are consistent with the underlying cosmology, with the exception of the very last redshift bin which, due to the lack of galaxies at the high-$z$ end of the distribution, lies $1\sigma$ above unity.
Likewise, we find that the amplitude of $b\sigma_8/\sigma_{8,\,\rm fid}$ remains rather flat and close to values of $\sim2$ up to $z\sim0.6$, after which it increases up to $\sim2.5$. We have checked this is a consequence of the trend followed by the large-scale bias of the sample implemented in the MD-PATCHY mocks, explaining why ADF $b\sigma_8/\sigma_{8,\,\rm fid}$ estimates also follow the same trend.

We also stress that, unlike ARF, ADF provide significantly biased estimates for $Ef\sigma_8$, which is likely due to its higher sensitivity to non-linearities. The bias on $Ef\sigma_8$ also impacts the ADF constraints on $b\sigma_8$ (both quantities are anti-correlated), which lie slightly above those from ARF. We hence discard the use of ADF in the remainder of this work.

Finally, we note that our small-scale, random-motion parameter $\sigma_v$ is largely unconstrained for most of our redshift shells, owing to the relatively large scales considered here. Modelling these random motions will however become necessary when probing smaller scales.

This preceding analysis allows us to confidently apply our methodology to the LOWZ+CMASS DR12 sample (see right panel of Fig.~\ref{fig:F2}). A first comparison with results from the MD-PATCHY mocks highlights the obvious similarity in the trend of $b\sigma_8/\sigma_{8,\,{\rm fid}}$ versus redshift. 
The $Ef\sigma_8$ estimates from LOWZ+CMASS DR12 sample are also consistent with the fiducial expectations, given the uncertainty in our measurements. The inferred amplitude of $Ef\sigma_8$ over its fiducial value lies 10\% above unity, ($A=1.10\pm 0.07$), and thus our measurement is 1.4$\sigma$ above {\it Planck}'s expectations. Uncertainties in the amplitude of $Ef\sigma_8$ grow at lower redshifts, for which data seems to hint some excess above the fiducial expectation. This may be due to a random fluctuation, but one could also speculate about late-time, non-linear physics entering differently as predicted by the MD-PATCHY mocks. 
So far we have discussed results for $R_{\rm NL} = 60~\hMpc$.
In Table~\ref{tab:tab1} we explore the dependence of our constraints on $R_{\rm NL}$ for slightly different values of this parameter. As we include smaller scales, the constraining power of our method increases -- uncertainties decrease from 8 to 6\% as we decrease $R_{\rm NL}$ from $70$ to $50~\hMpc$. We also observe there is a slight trend for $Ef\sigma_8$ ($b\sigma_8$) to decrease (increase) with decreasing $R_{\rm NL}$, as one would expect when non-linear power on transverse scales enters the analysis. In all cases, however, the constraints on our parameter combinations remain statistically unbiased, confirming the robustness of our analysis. In contrast, traditional ADF appear biased on all scales.

\begin{table}
\caption{\label{tab:tab1} Ratio of the amplitude of $Ef\sigma_8$ over its fiducial expectation for both ADF/ARF probes, COLA mocks, MD-PATCHY mocks, BOSS DR12 data, and three different $R_{\rm NL}$ choices.}
\begin{tabular}{ccccc}
Probe & $R_{\rm NL} [\hMpc]$ & COLA & PATCHY & BOSS DR12\\
\hline
    & $50$ & $1.07\pm 0.03$ & $0.96\pm 0.06$ & $1.06\pm 0.06$ \\
ARF & $60$ & $1.09\pm 0.04$ & $1.03\pm 0.07$ & $1.10\pm 0.07$ \\
    & $70 $ &$1.12\pm 0.04$ & $1.05\pm 0.08$ & $1.16\pm 0.08$ \\
\hline
    & $50$ & $1.13\pm 0.03$ & $0.19\pm 0.05$ & $0.29\pm 0.05$ \\
ADF & $60$ & $1.12\pm 0.04$ & $0.27\pm 0.06$ & $0.31\pm 0.06$ \\
    & $70$ & $1.15\pm 0.04$ & $0.35\pm 0.08$ & $0.44\pm 0.08$ \\
\end{tabular}
\end{table}

\begin{figure*}
\includegraphics[width=0.4\paperwidth]{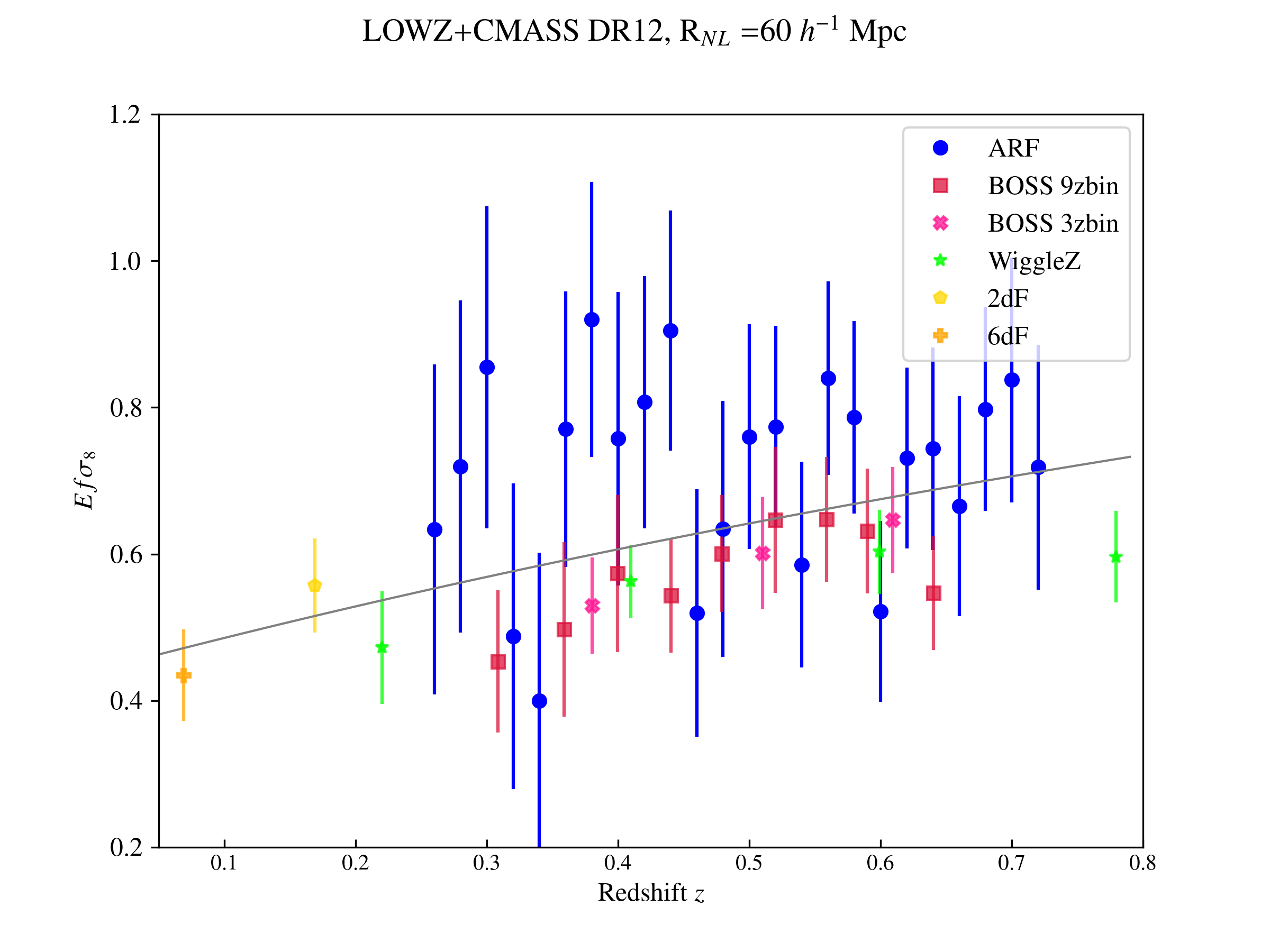}
\includegraphics[width=0.4\paperwidth]{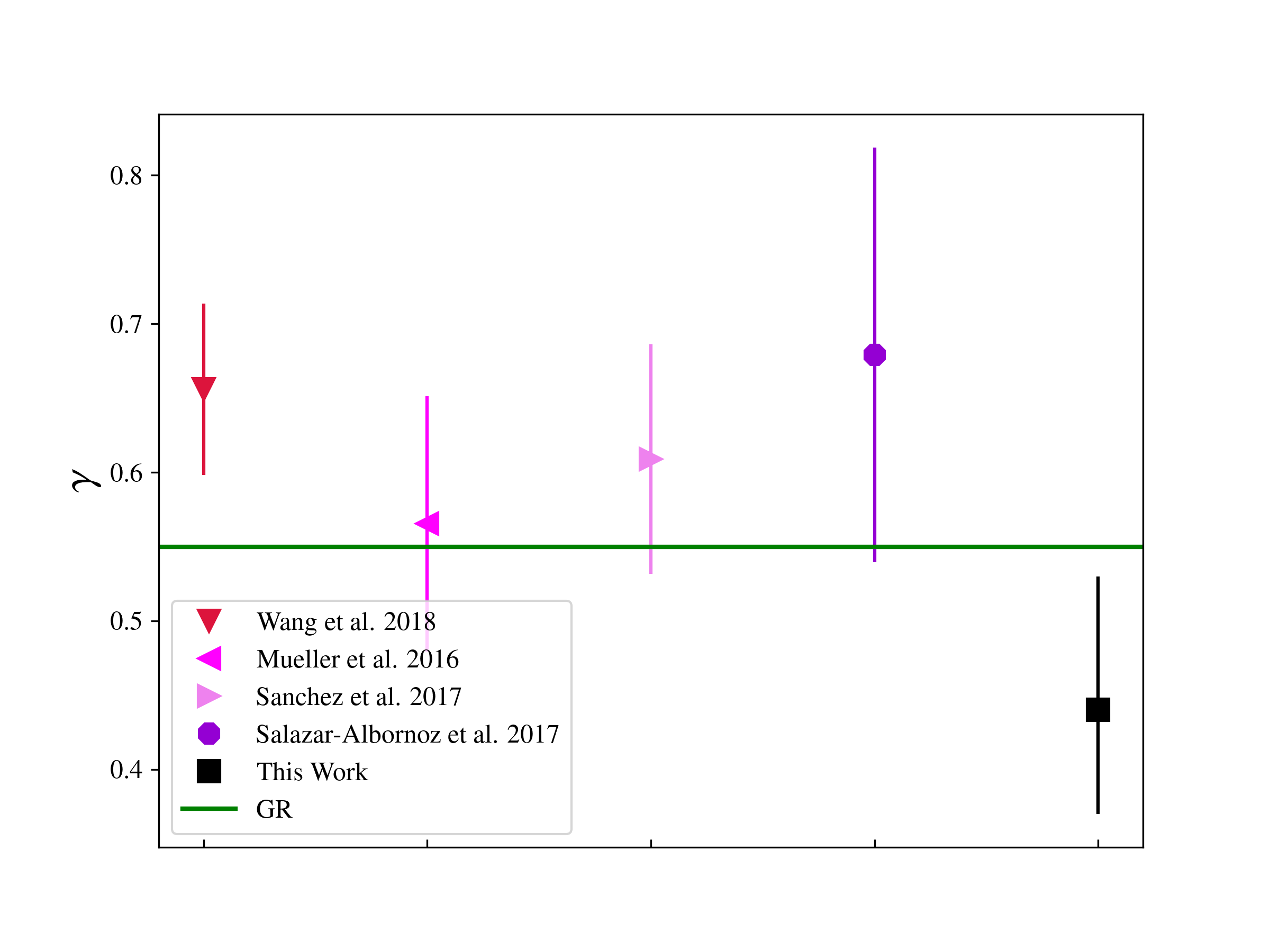}
\caption{
{\it (Left panel:)} Constraints on $Ef\sigma_8$ from ARF in LOWZ+CMASS DR12 (blue circles) in comparison with traditional $f\sigma_8$ constraints from recent spectroscopic surveys. The latter have been re-scaled by the value of $E$ predicted by our {\it Planck} fiducial model. Estimates from BOSS 9/3-$z$ bins \citep{wangetal18}, WiggleZ \citep{wigglez}, 2dF \citep{2dF}, and 6dF \citep{6dFGRs} were obtained using the anisotropic redshift space clustering. 
{\it (Right panel:)} Constraints on the $\gamma$ parameter obtained after combining {\it Planck} DR3 base $\Lambda$CDM model parameter chains with our measurements of $Ef\sigma_8(z)$ at $R_{\rm NL}=60\,\hMpc$. Our constraints are compatible (at $\sim1.4\sigma$) with the expectation from General Relativity ($\gamma_{\rm GR}\simeq 0.55$), but fall at the opposite side of most LOWZ+CMASS previous measurements based upon clustering.
}
\label{fig:F3}
\end{figure*}

Figure~\ref{fig:F2} and Table~\ref{tab:tab1} thus comprise one of the main results of this {\it letter}: the first tomographic measurement of the parameter $Ef\sigma_8$, at the $\sim 6-8$~\% level. This value is compatible to (although slightly higher than) the fiducial $\Lambda$CDM expectation. The parameter combination $Ef\sigma_8$ is obtained, for first time, from measurements of ARF, and is a direct effect of the action of gravity on the largest cosmological scales sampled by this galaxy survey. This same analysis of ARF provides the most precise tomographic measurement of the LOWZ+CMASS DR12 halo bias in the same redshift range. 

To put our work into context, in the left panel of Fig.~\ref{fig:F3} we compare our measurements to others obtained in the literature from the analysis of redshift-space distortions in anisotropic galaxy clustering. Explicitly, we consider a recent analysis of BOSS DR12 \cite{wangetal18}, WiggleZ \cite{wigglez}, 2dF \cite{2dF}, and 6dF \cite{6dFGRs} surveys. Since these constrain the parameter combination $f\sigma_8$, we have scaled these constraints by the value of $E(z)$ predicted by our {\it Planck} fiducial model.

First of all, we can clearly see the tomographic nature of our approach and that it can deliver constraints on $Ef\sigma_8$ at a similar level of relative accuracy as those on $f\sigma_8$ from traditional clustering approaches. Naturally, as we consider smaller subsets of the data, at each redshift our measurements are significantly noisier than if combined in wider redshift bins. In general, we see that previous analyses of LOW+CMASS DR12 and WiggleZ tend to prefer values smaller than that expected from {\it Planck}, contrary to our findings. We highlight that our analysis is restricted to large scales, in contrast with, e.g. the BOSS 9zbins \cite{wangetal18}, which employs separations $\sim$2.4 times smaller in which the modelling of non-linearities is significantly more important. 

Finally, in the right panel of Fig.~\ref{fig:F3}, we explore the implications of our measurements on the gravitational growth index $\gamma$, defined via $f \equiv d\log D_{\delta}/d\log a \simeq \Omega_m^{\gamma} (z)$. The combination of our measurements of $Ef\sigma_8(z)$ with \Planck~DR3 constraints in the minimal $\Lambda$CDM scenario allows us measuring $\gamma$ with $\lesssim 20~$\% uncertainty (see Fig.~\ref{fig:F3}): $\gamma=0.44^{+0.09}_{-0.07}$, which lies within $2$--$\sigma$ from the expectation in General Relativity ($\gamma_{\rm GR}\simeq 0.55$) 

{\it Discussion and conclusions:} This work demonstrates that ARF are a competitive tool for cosmological analysis, featuring several advantages over traditional approaches. We show that strong cosmological constraints are possible, even when restricting to large scales and simple modelings; that analysis can be performed without assuming any redshift-to-distance relation for a given fiducial cosmology; and with very little sensitivity to observational systematic errors.

Using ARF measured over 20 redshift bins in the LOWZ+CMASS DR12 samples, we constrain, for the first time, the action of gravity through $Ef\sigma_8$ at the 7~\% level, and $b\sigma_8$ with a $\lesssim 5~\%$ precision per redshift bin. By applying our analysis pipeline to two different sets of mocks, we demonstrated the robustness and accuracy of our approach. We also found that ARF contains additional information to that encoded in ADF, and that ADF are much more affected by non-linearities (thus we discarded them in our analysis). 
Interestingly, our ARF constraint on $\gamma$ points towards a {\em strong} effect of gravity ($\gamma<\gamma_{\rm GR}$), contrary to previous $\gamma$ measurements based on traditional clustering analyses down to small scales, which hint {\em weak} gravity ($\gamma>\gamma_{\rm GR}$) \cite{salazar_albornoz17,sanchezetal17,gilmarin17,wangetal18,Muelleretal18}. The tomographic character of our ARF study provides a highly competitive precision to our $\gamma$ estimate, only slightly surpassed by the (also tomographic) study of \cite{wangetal18}, which probes smaller (and potentially more non-linear) scales (down to $R=25~\hMpc$).


In the future, further developments of ARF modelling will enable us to exploit smaller cosmological scales and include relativistic effects and the effect of other cosmological parameters such as primordial non-Gaussianity. Furthermore, it will also be possible to generalise our formalism to analyse HI 21~cm or Lyman-$\alpha$ surveys, as well as other ongoing and upcoming galaxy/quasar surveys. 

{\it Acknowledgments}
 We thank S. Rodr\'\i guez-Torres and F.-S. Kitaura for help with the MD-PATCHY mocks. We acknowledge useful discussions with R. Adam, G. Hurier, C.~L\'opez San Juan, and L. Legrand. The authors acknowledge support from the Spanish Ministry of Science, Innovation, and Universities through the projects AYA2015-66211-C2-2 and PGC2018-097585-B-C2, and from the EU through projects PCIG9-GA-2011-294183 and ERC-StG/716151. We acknowledge the use of {\it Planck} and SDSS-III data.
{\it Data availability}
Specific data produced in this analysis available upon request. BOSS DR12 LOWZ and CMASS galaxy catalogues are accessible at \url{https://data.sdss.org/sas/dr12/boss/lss/} \citep{sdss_III}.

\bibliographystyle{mnras}
\bibliography{refs1.bib}

\begin{thebibliography}{}
\makeatletter
\relax
\def\mn@urlcharsother{\let\do\@makeother \do\$\do\&\do\#\do\^\do\_\do\%\do\~}
\def\mn@doi{\begingroup\mn@urlcharsother \@ifnextchar [ {\mn@doi@}
  {\mn@doi@[]}}
\def\mn@doi@[#1]#2{\def\@tempa{#1}\ifx\@tempa\@empty \href
  {http://dx.doi.org/#2} {doi:#2}\else \href {http://dx.doi.org/#2} {#1}\fi
  \endgroup}
\def\mn@eprint#1#2{\mn@eprint@#1:#2::\@nil}
\def\mn@eprint@arXiv#1{\href {http://arxiv.org/abs/#1} {{\tt arXiv:#1}}}
\def\mn@eprint@dblp#1{\href {http://dblp.uni-trier.de/rec/bibtex/#1.xml}
  {dblp:#1}}
\def\mn@eprint@#1:#2:#3:#4\@nil{\def\@tempa {#1}\def\@tempb {#2}\def\@tempc
  {#3}\ifx \@tempc \@empty \let \@tempc \@tempb \let \@tempb \@tempa \fi \ifx
  \@tempb \@empty \def\@tempb {arXiv}\fi \@ifundefined
  {mn@eprint@\@tempb}{\@tempb:\@tempc}{\expandafter \expandafter \csname
  mn@eprint@\@tempb\endcsname \expandafter{\@tempc}}}

\bibitem[\protect\citeauthoryear{{Abbott} et~al.,}{{Abbott}
  et~al.}{2018}]{DES1yr_GC_WL}
{Abbott} T.~M.~C.,  et~al., 2018, \mn@doi [\prd] {10.1103/PhysRevD.98.043526},
  \href {https://ui.adsabs.harvard.edu/abs/2018PhRvD..98d3526A} {98, 043526}

\bibitem[\protect\citeauthoryear{{Alam} et~al.,}{{Alam}
  et~al.}{2015}]{sdss_III}
{Alam} S.,  et~al., 2015, \mn@doi [\apjs] {10.1088/0067-0049/219/1/12}, \href
  {https://ui.adsabs.harvard.edu/abs/2015ApJS..219...12A} {219, 12}

\bibitem[\protect\citeauthoryear{Alam et~al.,}{Alam et~al.}{2017}]{Alam_2017}
Alam S.,  et~al., 2017, \mn@doi [Monthly Notices of the Royal Astronomical
  Society] {10.1093/mnras/stx721}, 470, 2617–2652

\bibitem[\protect\citeauthoryear{{Beutler} et~al.,}{{Beutler}
  et~al.}{2012}]{6dFGRs}
{Beutler} F.,  et~al., 2012, \mn@doi [\mnras]
  {10.1111/j.1365-2966.2012.21136.x}, \href
  {https://ui.adsabs.harvard.edu/abs/2012MNRAS.423.3430B} {423, 3430}

\bibitem[\protect\citeauthoryear{{Beutler} et~al.,}{{Beutler}
  et~al.}{2017a}]{Beutler2017a}
{Beutler} F.,  et~al., 2017a, \mn@doi [\mnras] {10.1093/mnras/stw2373}, \href
  {https://ui.adsabs.harvard.edu/abs/2017MNRAS.464.3409B} {464, 3409}

\bibitem[\protect\citeauthoryear{{Beutler} et~al.,}{{Beutler}
  et~al.}{2017b}]{Beutler2017b}
{Beutler} F.,  et~al., 2017b, \mn@doi [\mnras] {10.1093/mnras/stw3298}, \href
  {https://ui.adsabs.harvard.edu/abs/2017MNRAS.466.2242B} {466, 2242}

\bibitem[\protect\citeauthoryear{{Blake} et~al.,}{{Blake}
  et~al.}{2011}]{wigglez}
{Blake} C.,  et~al., 2011, \mn@doi [\mnras] {10.1111/j.1365-2966.2011.18903.x},
  \href {https://ui.adsabs.harvard.edu/abs/2011MNRAS.415.2876B} {415, 2876}

\bibitem[\protect\citeauthoryear{{Chaves-Montero}, {Hernandez-Monteagudo},
  {Angulo}  \& {Emberson}}{{Chaves-Montero} et~al.}{2019}]{Chaves-Montero_kSZ}
{Chaves-Montero} J.,  {Hernandez-Monteagudo} C.,  {Angulo} R.~E.,   {Emberson}
  J.~D.,  2019, arXiv e-prints, \href
  {https://ui.adsabs.harvard.edu/abs/2019arXiv191110690C} {p. arXiv:1911.10690}

\bibitem[\protect\citeauthoryear{Chuang et~al.,}{Chuang
  et~al.}{2017}]{Chuang_2017}
Chuang C.-H.,  et~al., 2017, \mn@doi [Monthly Notices of the Royal Astronomical
  Society] {10.1093/mnras/stx1641}, 471, 2370–2390

\bibitem[\protect\citeauthoryear{{Cuesta} et~al.,}{{Cuesta}
  et~al.}{2016}]{cuesta_bao_dr12}
{Cuesta} A.~J.,  et~al., 2016, \mn@doi [\mnras] {10.1093/mnras/stw066}, \href
  {https://ui.adsabs.harvard.edu/abs/2016MNRAS.457.1770C} {457, 1770}

\bibitem[\protect\citeauthoryear{Doré et~al.,}{Doré et~al.}{2014}]{sphereX}
Doré O.,  et~al., 2014, Cosmology with the SPHEREX All-Sky Spectral Survey
  (\mn@eprint {arXiv} {1412.4872})

\bibitem[\protect\citeauthoryear{{Galitzki}}{{Galitzki}}{2018}]{simons_obs_18}
{Galitzki} N.,  2018, arXiv e-prints, \href
  {https://ui.adsabs.harvard.edu/abs/2018arXiv181002465G} {p. arXiv:1810.02465}

\bibitem[\protect\citeauthoryear{{Gil-Mar{\'\i}n}, {Percival}, {Verde},
  {Brownstein}, {Chuang}, {Kitaura}, {Rodr{\'\i}guez-Torres}  \&
  {Olmstead}}{{Gil-Mar{\'\i}n} et~al.}{2017}]{gilmarin17}
{Gil-Mar{\'\i}n} H.,  {Percival} W.~J.,  {Verde} L.,  {Brownstein} J.~R.,
  {Chuang} C.-H.,  {Kitaura} F.-S.,  {Rodr{\'\i}guez-Torres} S.~A.,
  {Olmstead} M.~D.,  2017, \mn@doi [\mnras] {10.1093/mnras/stw2679}, \href
  {https://ui.adsabs.harvard.edu/abs/2017MNRAS.465.1757G} {465, 1757}

\bibitem[\protect\citeauthoryear{{Hernandez-Monteagudo}, {Chaves-Montero}  \&
  {Angulo}}{{Hernandez-Monteagudo} et~al.}{2019}]{arf_lett1}
{Hernandez-Monteagudo} C.,  {Chaves-Montero} J.,   {Angulo} R.~E.,  2019, arXiv
  e-prints, \href {https://ui.adsabs.harvard.edu/abs/2019arXiv191112056H} {p.
  arXiv:1911.12056}

\bibitem[\protect\citeauthoryear{{Hivon}, {G{\'o}rski}, {Netterfield}, {Crill},
  {Prunet}  \& {Hansen}}{{Hivon} et~al.}{2002}]{master}
{Hivon} E.,  {G{\'o}rski} K.~M.,  {Netterfield} C.~B.,  {Crill} B.~P.,
  {Prunet} S.,   {Hansen} F.,  2002, \mn@doi [\apj] {10.1086/338126}, \href
  {https://ui.adsabs.harvard.edu/abs/2002ApJ...567....2H} {567, 2}

\bibitem[\protect\citeauthoryear{{Kitaura} et~al.,}{{Kitaura}
  et~al.}{2016}]{patchy}
{Kitaura} F.-S.,  et~al., 2016, \mn@doi [\mnras] {10.1093/mnras/stv2826}, \href
  {https://ui.adsabs.harvard.edu/abs/2016MNRAS.456.4156K} {456, 4156}

\bibitem[\protect\citeauthoryear{{Laurent} et~al.,}{{Laurent}
  et~al.}{2017}]{Laurent17}
{Laurent} P.,  et~al., 2017, \mn@doi [\jcap] {10.1088/1475-7516/2017/07/017},
  \href {https://ui.adsabs.harvard.edu/abs/2017JCAP...07..017L} {2017, 017}

\bibitem[\protect\citeauthoryear{Levi et~al.,}{Levi et~al.}{2019}]{desi_2019}
Levi M.~E.,  et~al., 2019, The Dark Energy Spectroscopic Instrument (DESI)
  (\mn@eprint {arXiv} {1907.10688})

\bibitem[\protect\citeauthoryear{{Loureiro} et~al.,}{{Loureiro}
  et~al.}{2019}]{bossdr12_cl_tomography}
{Loureiro} A.,  et~al., 2019, \mn@doi [\mnras] {10.1093/mnras/stz191}, \href
  {https://ui.adsabs.harvard.edu/abs/2019MNRAS.485..326L} {485, 326}

\bibitem[\protect\citeauthoryear{{Mueller}, {Percival}, {Linder}, {Alam},
  {Zhao}, {S{\'a}nchez}, {Beutler}  \& {Brinkmann}}{{Mueller}
  et~al.}{2018}]{Muelleretal18}
{Mueller} E.-M.,  {Percival} W.,  {Linder} E.,  {Alam} S.,  {Zhao} G.-B.,
  {S{\'a}nchez} A.~G.,  {Beutler} F.,   {Brinkmann} J.,  2018, \mn@doi [\mnras]
  {10.1093/mnras/stx3232}, \href
  {https://ui.adsabs.harvard.edu/abs/2018MNRAS.475.2122M} {475, 2122}

\bibitem[\protect\citeauthoryear{{Percival} et~al.,}{{Percival}
  et~al.}{2004}]{2dF}
{Percival} W.~J.,  et~al., 2004, \mn@doi [\mnras]
  {10.1111/j.1365-2966.2004.08146.x}, \href
  {https://ui.adsabs.harvard.edu/abs/2004MNRAS.353.1201P} {353, 1201}

\bibitem[\protect\citeauthoryear{{Planck Collaboration} et~al.,}{{Planck
  Collaboration} et~al.}{2016}]{planck_parameters_15}
{Planck Collaboration} et~al., 2016, \mn@doi [\aap]
  {10.1051/0004-6361/201525830}, \href
  {http://adsabs.harvard.edu/abs/2016A%26A...594A..13P} {594, A13}

\bibitem[\protect\citeauthoryear{Ross et~al.,}{Ross et~al.}{2011}]{Ross_2011}
Ross A.~J.,  et~al., 2011, \mn@doi [Monthly Notices of the Royal Astronomical
  Society] {10.1111/j.1365-2966.2011.19351.x}, 417, 1350–1373

\bibitem[\protect\citeauthoryear{Ross et~al.,}{Ross et~al.}{2012}]{Ross_2012}
Ross A.~J.,  et~al., 2012, \mn@doi [Monthly Notices of the Royal Astronomical
  Society] {10.1111/j.1365-2966.2012.21235.x}, 424, 564–590

\bibitem[\protect\citeauthoryear{{Salazar-Albornoz} et~al.,}{{Salazar-Albornoz}
  et~al.}{2017}]{salazar_albornoz17}
{Salazar-Albornoz} S.,  et~al., 2017, \mn@doi [\mnras] {10.1093/mnras/stx633},
  \href {https://ui.adsabs.harvard.edu/abs/2017MNRAS.468.2938S} {468, 2938}

\bibitem[\protect\citeauthoryear{{S{\'a}nchez} et~al.,}{{S{\'a}nchez}
  et~al.}{2017}]{sanchezetal17}
{S{\'a}nchez} A.~G.,  et~al., 2017, \mn@doi [\mnras] {10.1093/mnras/stw2443},
  \href {https://ui.adsabs.harvard.edu/abs/2017MNRAS.464.1640S} {464, 1640}

\bibitem[\protect\citeauthoryear{Scaramella et~al.,}{Scaramella
  et~al.}{2014}]{Euclid_scaramella14}
Scaramella R.,  et~al., 2014, \mn@doi [Proceedings of the International
  Astronomical Union] {10.1017/s1743921314011089}, 10, 375–378

\bibitem[\protect\citeauthoryear{{Shafer} \& {Huterer}}{{Shafer} \&
  {Huterer}}{2015}]{shafer2015}
{Shafer} D.~L.,  {Huterer} D.,  2015, \mn@doi [\mnras] {10.1093/mnras/stu2640},
  \href {https://ui.adsabs.harvard.edu/abs/2015MNRAS.447.2961S} {447, 2961}

\bibitem[\protect\citeauthoryear{{Swanson}, {Tegmark}, {Hamilton}  \&
  {Hill}}{{Swanson} et~al.}{2008}]{mangle}
{Swanson} M.~E.~C.,  {Tegmark} M.,  {Hamilton} A. J.~S.,   {Hill} J.~C.,  2008,
  \mn@doi [\mnras] {10.1111/j.1365-2966.2008.13296.x}, \href
  {https://ui.adsabs.harvard.edu/abs/2008MNRAS.387.1391S} {387, 1391}

\bibitem[\protect\citeauthoryear{{Tassev}, {Zaldarriaga}  \&
  {Eisenstein}}{{Tassev} et~al.}{2013}]{colaSIMs}
{Tassev} S.,  {Zaldarriaga} M.,   {Eisenstein} D.~J.,  2013, \mn@doi [\jcap]
  {10.1088/1475-7516/2013/06/036}, \href
  {http://adsabs.harvard.edu/abs/2013JCAP...06..036T} {6, 036}

\bibitem[\protect\citeauthoryear{{Thornton} et~al.,}{{Thornton}
  et~al.}{2016}]{actpol_16}
{Thornton} R.~J.,  et~al., 2016, \mn@doi [\apjs] {10.3847/1538-4365/227/2/21},
  \href {https://ui.adsabs.harvard.edu/abs/2016ApJS..227...21T} {227, 21}

\bibitem[\protect\citeauthoryear{{Troxel} et~al.,}{{Troxel}
  et~al.}{2018}]{DES1yr_shear}
{Troxel} M.~A.,  et~al., 2018, \mn@doi [\prd] {10.1103/PhysRevD.98.043528},
  \href {https://ui.adsabs.harvard.edu/abs/2018PhRvD..98d3528T} {98, 043528}

\bibitem[\protect\citeauthoryear{{Wang}, {Zhao}, {Chuang}, {Pellejero-Ibanez},
  {Zhao}, {Kitaura}  \& {Rodriguez-Torres}}{{Wang} et~al.}{2018}]{wangetal18}
{Wang} Y.,  {Zhao} G.-B.,  {Chuang} C.-H.,  {Pellejero-Ibanez} M.,  {Zhao} C.,
  {Kitaura} F.-S.,   {Rodriguez-Torres} S.,  2018, \mn@doi [\mnras]
  {10.1093/mnras/sty2449}, \href
  {https://ui.adsabs.harvard.edu/abs/2018MNRAS.481.3160W} {481, 3160}

\makeatother
\end{thebibliography}
\label{lastpage}


%

\end{document}